\documentclass[reprint,amsmath,amssymb, aps]{revtex4-1}

\usepackage{graphicx}% Include figure files
\usepackage{dcolumn}% Align table columns on decimal point
\usepackage{bm}% bold math
\begin{document}

\title{Magnetotransport properties of the triply degenerate node topological semimetal: tungsten carbide}

\author{J. B. He$^{1,2}$}
\thanks{These authors contributed equally to this work.}
\author{D. Chen$^{1}$}
\thanks{These authors contributed equally to this work.}
\author{W. L. Zhu$^{1,3}$}
\author{S. Zhang$^{1}$}
\author{L. X. Zhao$^{1,3}$}
\author{Z. A. Ren$^{1,4}$}
\author{G. F. Chen$^{1,3,4}$}
\email{gfchen@iphy.ac.cn}

\affiliation{
$^1$Institute of Physics and Beijing National Laboratory for Condensed Matter Physics, Chinese Academy of Sciences, Beijing 100190, China\\
$^2$College of Physics and Electronic Engineering, Nanyang Normal University, Nanyang 473061, China\\
$^3$School of Physical Sciences, University of Chinese Academy of Sciences, Beijing 100190, China\\
$^4$Collaborative Innovation Center of Quantum Matter, Beijing 100190, China\\
}
\date{\today}
\begin{abstract}

We report the magnetoresistance (MR), Hall effect, and de Haas-van Alphen (dHvA) effect studies of single crystals of tungsten carbide, WC, which is predicted to be a new type of topological semimetal with triply degenerate nodes. With the magnetic field rotated in the plane perpendicular to the current, WC shows a field induced metal-to-insulator-like transition and large nonsaturating quadratic MR at low temperatures. As the magnetic field parallel to the current, a pronounced negative longitudinal MR only can be observed for the certain direction of current flow. Hall effect indicates WC is a perfect compensated semimetal, which may be related to the large nonsaturating quadratic MR. The analysis of dHvA oscillations reveals that WC is a multiband system with small cross-sectional areas of Fermi surface and light cyclotron effective masses. Our results indicate that WC is an ideal platform to study the recently proposed ``New Fermions'' with triply degenerate crossing points.

\end{abstract}

\maketitle

The recently discovered topological semimetals (TSMs) have been studied intensively in condensed matter physics and material science due to
the topologically protected band structure, which exhibits some exotic physical properties such as extremely large magnetoresistance (MR), ultrahigh carrier mobility, anomalous Berry phase, and/or negative longitudinal MR \cite{TSM_W,TSM_RMP1,TSM_RMP2}. Generally, TSMs can be classified into several categories according to the degeneracy and momentum space distribution of the nodal points. For example, in Dirac semimetals (DSMs) \cite{Na3Bi_C,Na3Bi_ARPES,Na3Bi_MR,Cd3As2_C,Cd3As2_ARPES,Cd3As2_MR} and Weyl semimetals (WSMs) \cite{TaAs_Science,TaAs_PRX_Weng,TaAs_PRX_Ding,TaAs_PRX_Chen}, two double- and non-degeneracy bands cross near the Fermi level (E$_{F}$), which lead to the four-fold degenerate Dirac point and two-fold degenerate Weyl point, respectively. In contrast to discrete points in DSMs and WSMs, two bands cross each other along a line in the Brillouin zone in nodal line semimetals \cite{NLSM1,NLSM2}.

Besides the three kinds of TSMs mentioned above, some new types of TSMs with three-, six-, or eight-fold degenerate points near E$_{F}$ were proposed and these degenerate points appear at high-symmetry points in nonsymmorphic space groups \cite{TPSM_Science}. Based on a new mechanism, several materials with WC-type structure that belongs to symmorphic space groups, such as MoP, WC, TaN, and ZrTe et al., were then predicted to possess three-fold degenerate crossing points (TPs), formed by the crossing of a double-degeneracy band along a high-symmetry direction in momentum space and a non-degeneracy band \cite{TPSM_PRX,WC_Hasan,TaN_PRB,ZrTe_PRB}. However, so far, there is few experimental investigation about this kind of materials. Recently, angle-resolved photoemission spectroscopy (ARPES) measurements have confirmed the presence of TPs in MoP \cite{MoP_ARPES}. Unfortunately, in contrast to WC/TaN/ZrTe, the TPs of MoP are far away from E$_{F}$ \cite{WC_Hasan,TaN_PRB,ZrTe_PRB,MoP_ARPES}, and it is expected that there will be no fascinating transport property observed in this compound.

\begin{figure}
\includegraphics[width=9cm, height=6cm]{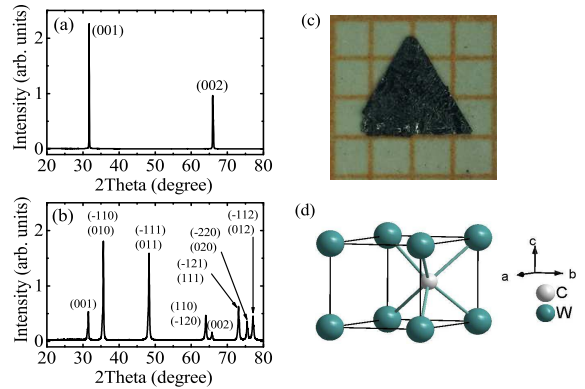}
\caption{\label{XRD}(Color online) (a) and (b) Single crystal and powder XRD pattern for WC, respectively. (c) The typical photography of the grown WC crystal. (d) The crystal structure of WC.
}
\end{figure}

 \begin{figure*}
\includegraphics[width=18cm, height=12cm]{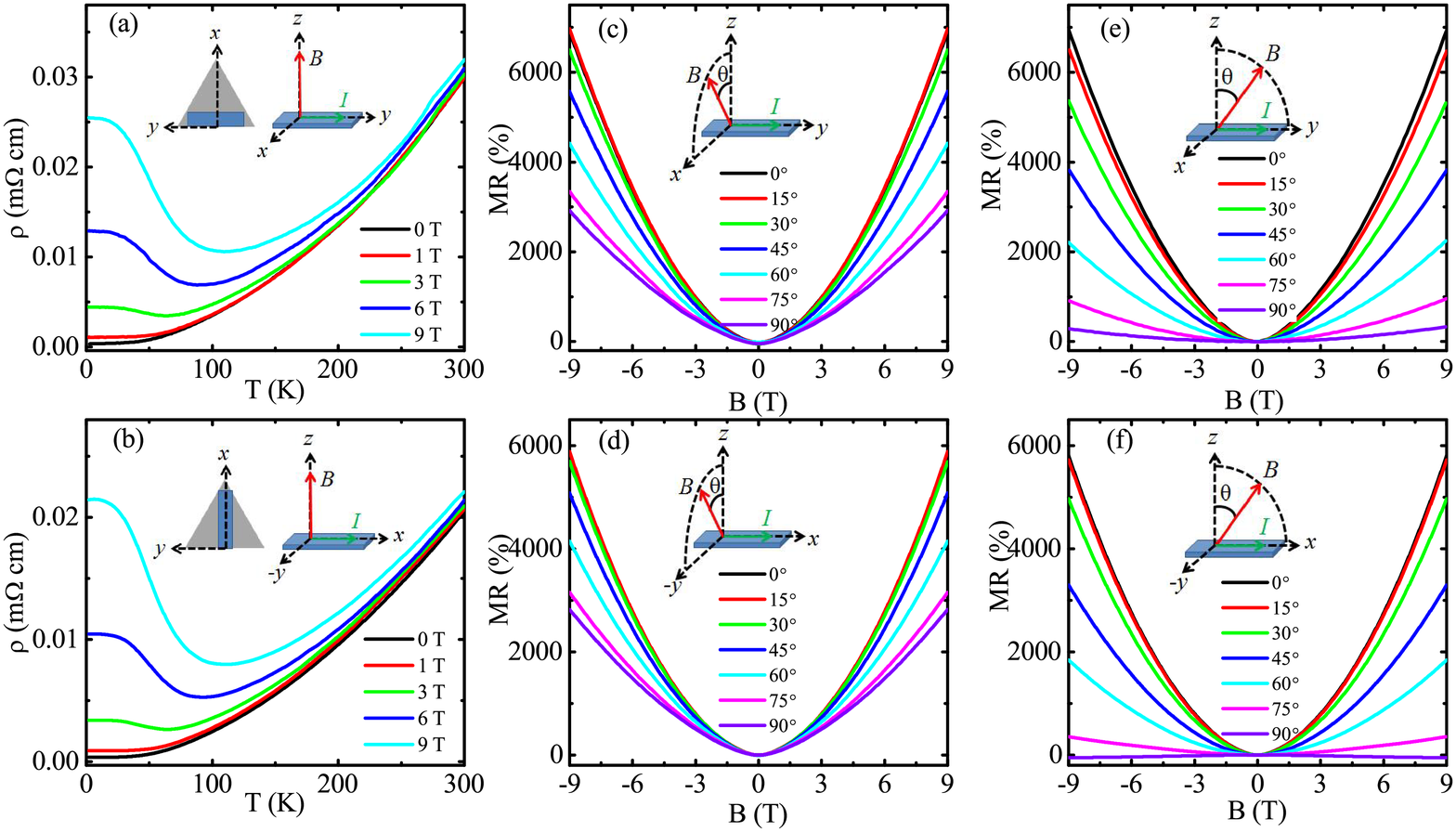}
\caption{\label{RT}(Color online) Transport properties of WC. The insets depict the corresponding measurement configurations. All measurements were performed on the bar specimens cut from the equilateral triangle shape single crystals. The electric current flows along the \emph{y}-axis (\emph{I} $\parallel$ \emph{y}) in (a), (c), and (e), and along the \emph{x}-axis (\emph{I} $\parallel$ \emph{x}) in (b), (d), and (f), respectively. (a) and (b) Temperature dependence of the resistivity $\rho$ under selected magnetic fields with \emph{B} $\parallel$ \emph{z} for both \emph{I} $\parallel$ \emph{y} and \emph{I} $\parallel$ \emph{x}, respectively. (c) and (d) Magnetic field dependence of MR at 2 K with magnetic field rotated in the plane perpendicular to the current for both \emph{I} $\parallel$ \emph{y} and \emph{I} $\parallel$ \emph{x}, respectively. (e) and (f) Magnetic field dependence of MR at 2 K with magnetic field tilted from \emph{B} $\perp$ \emph{I} ($\theta$ = 0$^\circ$) to \emph{B} $\parallel$ \emph{I} ($\theta$ = 90$^\circ$) for both \emph{I} $\parallel$ \emph{y} and \emph{I} $\parallel$ \emph{x}, respectively.
}
\end{figure*}

In this work, we have successfully grown the high quality single crystals of WC and carried out the MR, Hall, and de Haas-van Alphen (dHvA) effect studies. We found that WC is a perfect compensated semimetal, leading to a field induced metal-to-insulator-like transition and large nonsaturating MR as observed in other TSMs. Furthermore, a pronounced negative longitudinal MR for the certain current direction may be ascribed to Weyl nodes, which could split from triply-degenerate nodes in the presence of a Zeeman coupling.

High-quality single crystals of WC were grown by flux method. The starting materials, W, C, and Co, were mixed in the ratio of 1 : 1 : 5, and put into a graphite crucible with a cap. The crucible was heated to 1700 $^{\circ}$C, held for 50 h, and then cooled to 1400 $^{\circ}$C at a rate of 1 $^{\circ}$C/h in argon atmosphere. After the Co flux was removed by dissolving in a warm hydrochloric acid solution, the obtained single crystals are in the form of equilateral triangle shape with sides of 3 mm and thick of 1 mm, as shown in Fig. \ref{XRD}(c). The crystal structure was characterized by X-ray diffraction (XRD) using PANalytical diffractometer with Cu $K_{\alpha}$ radiation at room temperature. The elemental compositions were checked by Oxford X-Max energy dispersive x-ray (EDX) spectroscopy analysis in a Hitachi S-4800 scanning electron microscope. Electrical transport and magnetic measurements were performed using Quantum Design PPMS-9 T and MPMS-7 T SQUID VSM system, respectively.

The crystal structure of WC is illustrated in Fig. \ref{XRD}(d). WC crystalizes in simple hexagonal structure with a space group of \emph{P$\overline{6}$m2} (No. 187), which contains one formula unit in the unit cell \cite{WC_Structure}. W and C atoms occupy the 1\emph{a} (0, 0, 0) and 1\emph{d} (1/3, 2/3, 1/2) Wyckoff positions, respectively. Figure \ref{XRD} (a) shows the single crystal XRD pattern with (00\emph{l}) reflections. The powder XRD pattern, as shown in Fig. \ref{XRD} (b), can be well indexed to the structure of WC with the lattice constants \emph{a} = \emph{b} = 2.91 ${\AA}$ and \emph{c} = 2.84 ${\AA}$, which are comparable to the previously reported value \cite{WC_Structure}. The result of EDX indicates that the average W:C atomic ratio is close to 1:1 and there is no obvious Co impurity.

The temperature dependence of the resistivity under selected magnetic fields with \emph{B} $\parallel$ \emph{z} for both \emph{I} $\parallel$ \emph{y} and \emph{I} $\parallel$ \emph{x} are presented in Fig. \ref{RT}(a) and (b), respectively. In zero field, the resistivity shows metallic behavior with a large residual resistivity ratio [RRR = $\rho$(300 K)/$\rho$(2 K)] up to 85 and 60 for both current directions, indicating that the grown single crystal is of high quality and has few vacancies, as observed in EDX. When the magnetic fields were applied, a magnetic field induced metal-to-insulator-like transition and a resistivity plateau are present at low temperatures. The temperature corresponding to the minimum in the resistivity increases with increasing field. These phenomena have also been observed in the recently discovered semimetals such as WTe$_{2}$, TaAs, and LaSb \cite{WTe2,TaAs_PRX_Chen,LaSb}, although the mechanism is still unclear.

Figure \ref{RT} (c) and (d) show the magnetic field dependence of MR at 2 K with magnetic field rotated in the plane perpendicular to the current. When $\theta$ = 0$^\circ$ (\emph{B} $\parallel$ \emph{z}), the MR is maximized and can reach to about 7000\% and 6000\% at 9 T for both \emph{I} $\parallel$ \emph{y} and \emph{I} $\parallel$ \emph{x}, respectively. With increasing $\theta$ to 90$^\circ$ (\emph{B} $\parallel$ \emph{x} or \emph{B} $\parallel$ \emph{y}), MR gradually decreases and gets the minimum of 2900\% and 2800\% for both current directions. For all of the magnetic field directions, the dependence of MR on magnetic field is close to quadratic with no indication of saturation, which could be attributed to the perfect compensated of electrons and holes, similar to that observed in WTe$_{2}$ and $\alpha$-As \cite{WTe2_MR,As_PRB}.

\begin{figure}
\includegraphics[width=7cm, height=6cm]{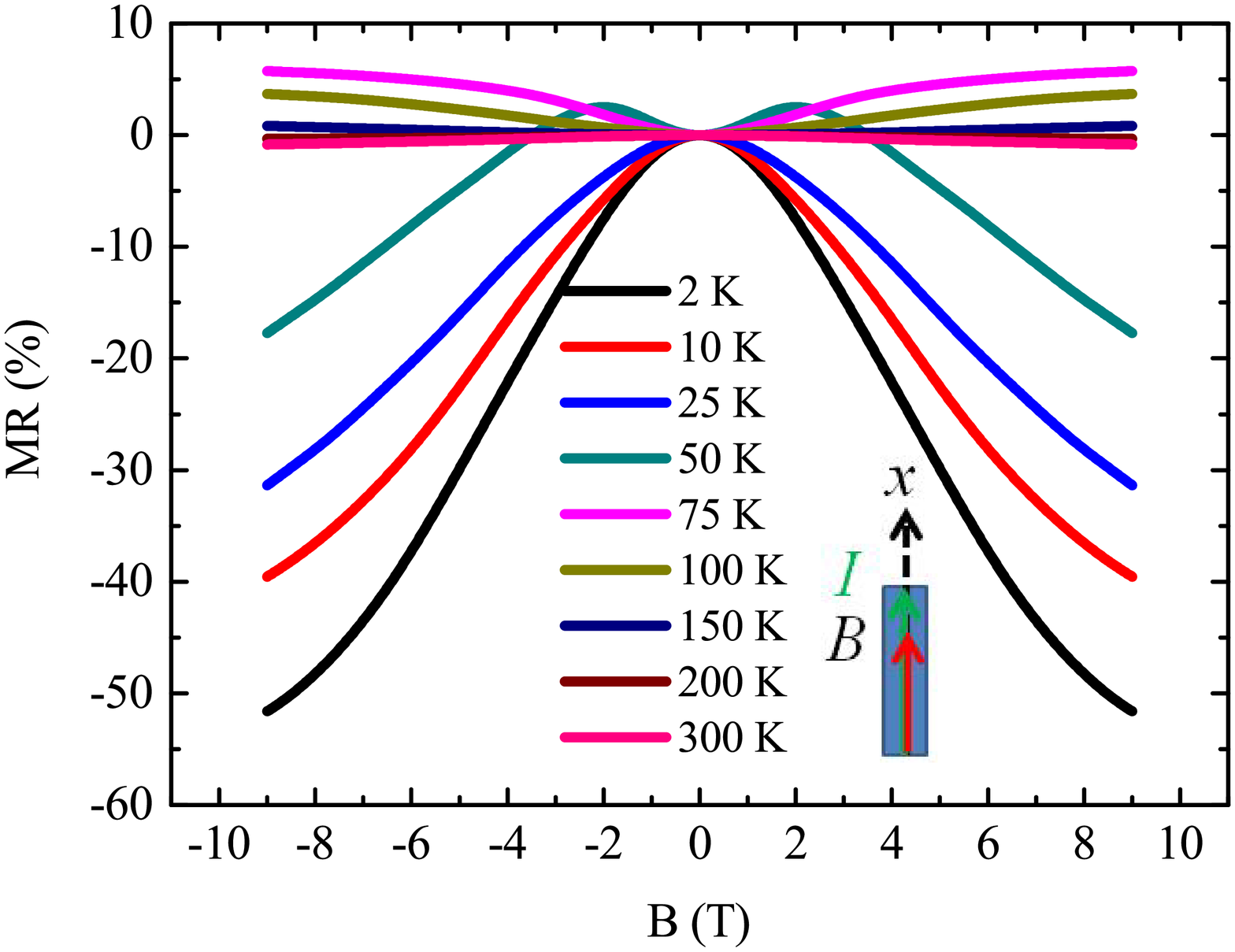}
\caption{\label{NMR}(Color online) Magnetic field dependence of MR at various temperatures for \emph{B} $\parallel$ \emph{I} $\parallel$ \emph{x}. The inset depicts the measurement configuration.
}
\end{figure}

The magnetic field dependence of MR with magnetic field tilted from \emph{B} $\perp$ \emph{I} ($\theta$ = 0$^\circ$) to \emph{B} $\parallel$ \emph{I} ($\theta$ = 90$^\circ$) at 2 K are shown in Fig. \ref{RT}(e) and (f) for both \emph{I} $\parallel$ \emph{y} and \emph{I} $\parallel$ \emph{x}, respectively. The MR is maximized for \emph{B} $\perp$ \emph{I} ($\theta$ = 0$^\circ$) and decreases with the $\theta$ increasing for both current directions. When \emph{B} $\parallel$ \emph{I} ($\theta$ = 90$^\circ$), the MR drop to a minimum of 300\% at \emph{B} = 9 T and T = 2 K for \emph{I} $\parallel$ \emph{y}, however, a negative MR with a value of about $-50\%$ was observed for \emph{I} $\parallel$ \emph{x} at the same condition. As shown in Fig. \ref{NMR}, the negative MR decreases with increasing temperature for \emph{B} $\parallel$ \emph{I} $\parallel$ \emph{x}. The theoretical calculation has pointed out that the magnetotransport properties depend on the direction of applied magnetic field and the triply-degenerate nodes could split into pairs of Weyl nodes with opposite chirality as the presence of a Zeeman coupling along the z direction in WC \cite{TPSM_PRX,WC_Hasan}. However, the negative MR is not directional in Na$_{3}$Bi, GdPtBi, and $\alpha$-As with the effect of Zeeman field \cite{Na3Bi_MR,GdPtBi,As_PRB}. Although the negative MR in WC is sensitive to the special crystal axis, as a possibility, we are reasonable to believe that the negative MR may be associated with the chiral anomaly of Weyl nodes, which may be split from triply-degenerate nodes in the presence of a Zeeman coupling at the certain direction. Further band structure calculations and ARPES measurements are necessary to clarify this issue.

 \begin{figure}
\includegraphics[width=9cm, height=6cm]{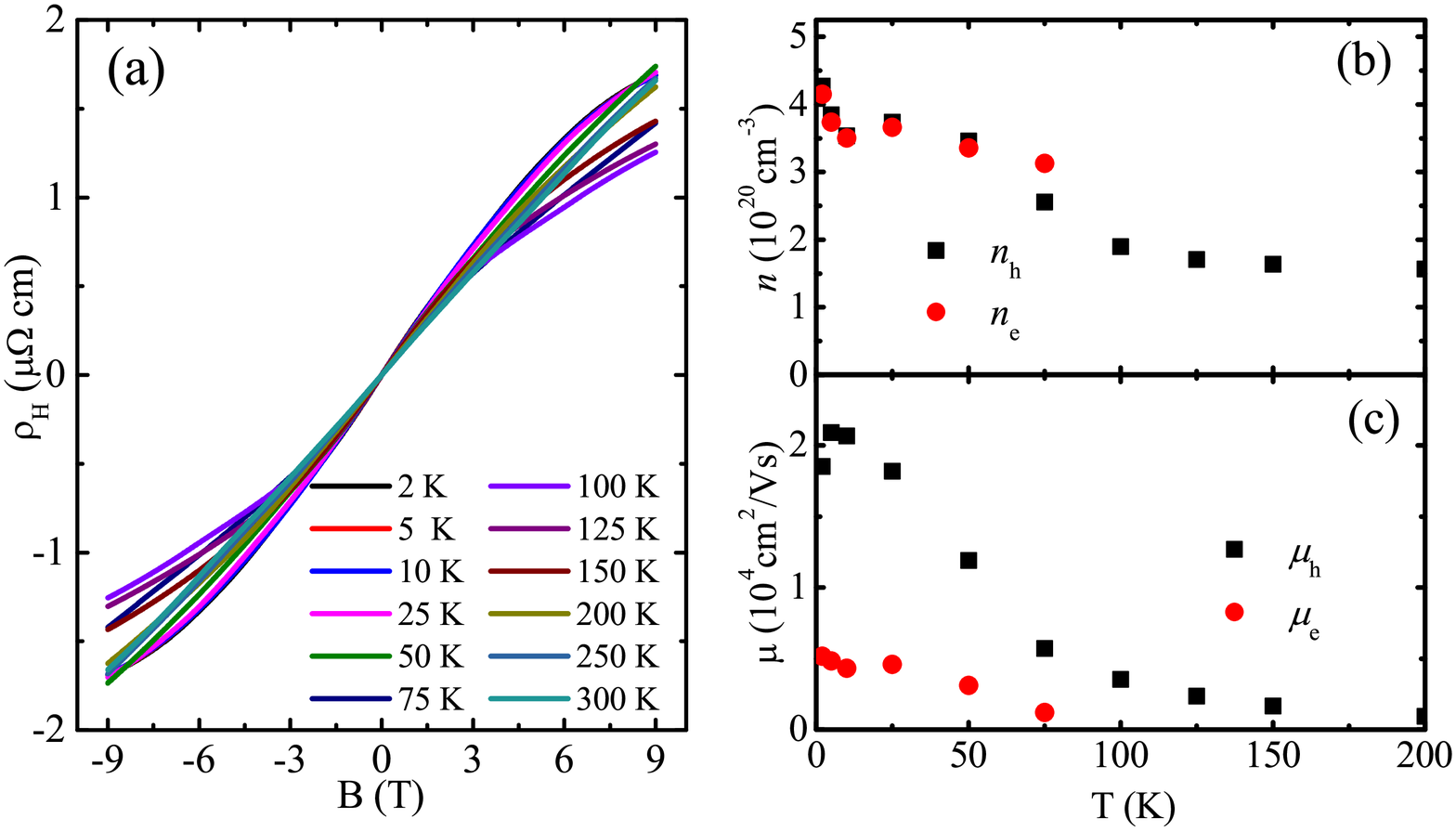}
\caption{\label{Hall}(Color online) (a) The magnetic field dependence of Hall resistivity $\rho$$_{H}$ at various temperatures. (b) and (c) The temperature dependence of carrier densities and carrier mobilities of electrons and holes obtained by fitting Hall conductivity $\sigma$$_{H}$ using the two- or single-band model.
}
\end{figure}

To investigate the mechanism of the large nonsaturating quadratic MR observed in WC, we measured the magnetic field dependence of Hall resistivity $\rho$$_{H}$ at various temperatures, as shown in Fig. \ref{Hall}(a). The positive Hall coefficient R$_{H}$ = $\rho$$_{H}$ /B shows that hole-type carriers are dominant in the transport process. Furthermore, the deviation from linear behavior of the Hall resistivity implies that WC is a multiband system. Thus, we analyzed the Hall conductivity $\sigma$$_{H}$ using the semi-classical two-band model:
\begin{equation}\label{1}
\sigma_{H} = (\frac{n_{h}\mu_{h}^{2}}{1+(\mu_{h}B)^2}-\frac{n_{e}\mu_{e}^{2}}{1+(\mu_{e}B)^2})eB,
\end{equation}
where $n_{h}$ ($n_{e}$) and $\mu_{h}$ ($\mu_{e}$) denote the carrier density and mobility of hole (electron), respectively. However, above 75 K, two-band model is failed and the single-band model is used. The temperature dependence of the carrier density and mobility extracted from the Hall conductivity fitting are shown in Fig. \ref{Hall}(b) and (c). The density and mobility of the two types of carriers increase with decreasing temperature. At low temperatures, the densities of holes and electrons are quite close, however, the mobility of holes is larger than that of electrons. This demonstrates that the large nonsaturating quadratic MR could stem from the perfect electron-hole compensation and high hole mobilities, similar to those observed in WTe$_{2}$ and $\alpha$-As \cite{WTe2_MR,As_PRB}.

 \begin{figure}
\includegraphics[width=9cm, height=8cm]{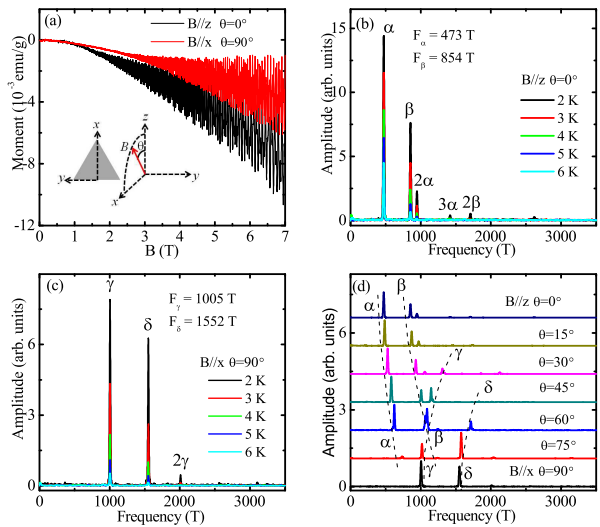}
\caption{\label{dHvA}(Color online)(a) Magnetization isotherms M vs magnetic field B for \emph{B} $\parallel$ \emph{z} ($\theta = 0^\circ$) and \emph{B} $\parallel$ \emph{x} ($\theta = 90^\circ$) at 2 K. The inset depicts the measurement configuration. (b) and (c) The FFT spectra of the corresponding dHvA oscillations at various temperatures for \emph{B} $\parallel$ \emph{z} and \emph{B} $\parallel$ \emph{x}. (d) The FFT spectra of the corresponding dHvA oscillations at 2 K rotating the field from \emph{B} $\parallel$ \emph{z} to \emph{B} $\parallel$ \emph{x}. The dashed lines are guides for eyes.
}
\end{figure}

DHvA effect is one of the most powerful tools to probe the properties of FS except for ARPES. As shown in Fig. \ref{dHvA}(a), we present the isothermal magnetization measured up to 7 T of a single crystal for both \emph{B} $\parallel$ \emph{z} and \emph{B} $\parallel$ \emph{x}, which exhibits obvious dHvA oscillations superimposed on a diamagnetic background. The fast Fourier transform (FFT) spectra of dHvA oscillations, as shown in Fig. \ref{dHvA}(b) and (c), yield several fundamental frequencies, F$_{\alpha}$ = 473 T and F$_{\beta}$ = 854 T for \emph{B} $\parallel$ \emph{z}, and F$_{\gamma}$ = 1005 T and F$_{\delta}$ = 1552 T for \emph{B} $\parallel$ \emph{x}, indicating WC is a multiband system, consistent with the result of Hall effect. Using the Onsager relation, \emph{F} = $\hbar$/(2$\pi$e)\emph{A$_{F}$}, where $\hbar$ is Planck's constant divided by 2$\pi$, e is the charge on the bare electron, and A$_{F}$ is the cross-sectional area of FS, it is calculated that the corresponding A$_{F}$ are 4.52, 8.15, 9.60, and 14.81 nm$^{-2}$, respectively, which are only about 0.80, 1.50, 1.74 and 2.68$\%$ of the first Brillouin zone. By fitting the thermal damping term \cite{dHvA}:
\begin{equation}\label{2}
\emph{R$_{T}$} = \frac{2\pi^{2}k_{B}Tm^{\ast}/{\hbar}eB}{sinh(2\pi^{2}k_{B}Tm^{\ast}/{\hbar}eB)} \propto Amp.
\end{equation}
where m* is the cyclotron effective mass, and k$_{B}$ is Boltzman's constant, the cyclotron effective masses are estimated to be 0.19, 0.31, 0.34, and 0.43 m$_{e}$ (m$_{e}$ is the bare electronic mass), respectively. The small cross-sectional areas of FS and the light cyclotron effective masses are the very important features of the topological semimetals. Figure \ref{dHvA}(d) shows the FFT spectra of the corresponding dHvA oscillations at 2 K rotating the field from \emph{B} $\parallel$ \emph{z} to \emph{B} $\parallel$ \emph{x} in the xz plane. As previously reported \cite{WC_dHvA}, the values of fundamental frequencies gradually change with the angle, indicating the low-frequency branches are anisotropic and may be the nearly ellipsoidal pockets.

In summary, we have successfully grown the high quality single crystals of WC, which is predicted to be a triply degenerate node topological semimetal candidate. It shows a metallic behavior at zero field and a field induced metal-to-insulator-like transition with magnetic field applied. The analysis of Hall effect and dHvA oscillations indicates WC is a perfect compensated system with small cross-sectional areas of FS and light cyclotron effective masses, leading to the large nonsaturating quadratic MR for \emph{B} $\perp$ \emph{I} at low temperatures. The pronounced negative longitudinal MR for the certain current direction may be caused by the Weyl nodes, split from triply-degenerate nodes in the presence of a Zeeman coupling. We hope this work can provide valuable clues for further research on WC and other new types of TSMs.

We would like to thank Xi Dai for helpful discussions. This work was supported by the National Basic Research Program of China 973 Program (Grant No. 2015CB921303), the National Key Research Program of China (Grant No. 2016YFA0300604), the Strategic Priority Research Program (B) of Chinese Academy of Sciences (Grant No. XDB07020100), and the Natural Science Foundation of China (Grant No. 11404175).

\bibliography{References}

\end{document}